\documentclass[twocolumn,prb,showkeys,amsfonts,showpacs]{revtex4}
\usepackage{amssymb}
\usepackage[dvips]{graphicx}
\usepackage{dcolumn}
\usepackage{amsmath}
\usepackage{bm}
\begin{document}
\title{Electric field and exciton structure in CdSe nanocrystals}
\author{E. Men\'{e}ndez-Proupin}
\email{eariel99@yahoo.com}
\altaffiliation[Present address: ]{Departamento de F\'\i sica, Universidad de Santiago de
Chile, Casilla 307, Santiago 2, Chile.}
\author{C. Trallero-Giner}
\affiliation{IMRE-Facultad de F\'{\i}sica, Universidad de La Habana, Vedado
10400, La Habana, Cuba}
\date{\today }
\begin{abstract}
Quantum Stark effect in semiconductor nanocrystals is theoretically
investigated, using the effective mass formalism within a $4\times 4$
Baldereschi-Lipari Hamiltonian model for the hole states. General
expressions are reported for the hole eigenfunctions at zero electric field.
Electron and hole single particle energies as  functions of the electric
field ($\mathbf{E}_{QD}$) are reported. Stark shift and binding energy of the
excitonic levels are obtained by full diagonalization of the correlated
electron-hole Hamiltonian in presence of the external field. Particularly, the
structure of the lower excitonic states and their symmetry properties in CdSe
nanocrystals are studied. It is found that the dependence of the exciton
binding energy upon the applied field is strongly reduced for small quantum
dot radius. Optical selection rules for absorption and luminescence are obtained.
The electric-field induced quenching of the optical spectra as a
function of $\mathbf{E}_{QD}$ is studied in terms of the exciton
dipole matrix element. It is predicted that photoluminescence spectra present
anomalous field dependence of the emission lines.
These results agree in
magnitude with experimental observation and with the main features of
photoluminescence experiments in nanostructures.
\end{abstract}

\maketitle

\section{Introduction}

Semiconductor nanostructures under longitudinal electric field produce
pronounced effects on optical properties.
\cite{yoffe,mendez,miller,kapteyn,miller88,raydmond,Fry}
 It has been shown that the
field induces a red shift of the exciton peaks in the photoluminescence and
electro-optical spectra. The shift of the excitonic peaks to lower energy
with the increasing electric fields is known as quantum Stark effect, while
the reduction of the overlapping between the electron-hole pair wave
function by the field is related to quenching of the fundamental transition
in the luminescence spectrum. Zero dimensional systems as colloidal
semiconductor quantum dots (QDs) under electric fields are appropriate
candidates for several device applications, including optical computing and
fiber-optical communication (see Ref.~\onlinecite{science97} and references
therein). Also, the micro-photoluminescence spectroscopic technique in
single spherical QDs has allowed to study fundamental issues of the excitonic
states.\cite{blanton}

The electro-optical properties, the Stark shift, and the dependence of the
peak intensity in the optical spectra upon the applied field
should depend strongly upon the details of the band structure. This
was demonstrated for quantum wells in the 80's.\cite{vina87,bauer87,bauer88b}
It is interesting to investigate the analogous effects in QD,
as the three-dimensional confinement causes properties that
are beyond the naive enhancement of the effects observed in quantum wells.
For example, in QDs the band dispersion
no longer exists, the energy spectrum is totally discrete and
depends qualitatively upon the dot geometry. Moreover, the high
surface to volume ratio originates effects that are intrinsic to QDs.
Perhaps, the most spectacular finding up to date is the discovery of
high luminescence in porous Si,\cite{canham90} where QDs are believed to play
an important role.\cite{wolkin99} Other striking effects can be found
in the dark magneto-exciton luminescence,\cite{rosen} and the blinking
and spectral shifting of single QD luminescence under external
electric fields.\cite{science97}

Calculations of the quantum Stark effect in spherical QDs in the strong
confinement regime have been performed in the framework of the parabolic model.
\cite{wen95,casadostark} The simple parabolic model was
able to provide a relative good picture
for the description of the electronic states in the conduction band. This
approximation breaks down for the calculation of the hole levels due to the
fourfold degeneracy and the admixture of the light and heavy-hole bands
present in the II-VI and II-III compounds with zinc-blende lattice
structure. A reliable description of the energy band dispersion is ruled by
the Baldereschi-Lipari Hamiltonian.\cite{bald} Within this approach,
calculations of the influence of an external electric
field $\mathbf{E}_{QD}$ were done in Ref.~\onlinecite{baixia98}.
However, this calculation presents several limitations (for a
detailed discussion see Ref.~\onlinecite{comment2chang98}).
As it is well known, the dependence of the
interband   optical transitions upon the light frequency (absorbed or emitted)
 reflects the  structure of the conduction and valence bands.

In this paper we study the excitonic Stark effect of spherical QDs taking
into account the valence band admixture using
the Baldereschi-Lipari Hamiltonian\cite{bald}.
Supported by a rigorous
treatment of the exciton wavefunctions, we have obtained
the interband dipole matrix elements taking into account the fundamental symmetry
properties of the QD.  From the dipole matrix elements  we have obtained
the optical selection rules that allow the identification
of the exciton levels observed in absorption and luminescence experiments.
We present numerical calculations that reveal
the combined effects of  band admixture, Coulomb interaction,
QD size and electric field intensity.

In Sec. II we examine the energy dependence upon $\mathbf{E}_{QD}$ for the
electrons and holes in CdSe QDs. Explicit analytical solutions for the hole
levels at $\mathbf{E}_{QD}=0$ are derived. The influence of  Coulomb
correlation and valence band coupling on the quantum Stark effect is
analyzed in Sec. III. Section IV is devoted to study the electric-field induced optical
properties.  The main results of the paper are summarized in Sec. V.

\section{Single particle states}

In the effective mass approximation, the Stark effect on the electronic states
at the bottom of the conduction band can be represented by products of the $%
\Gamma _{6}$ Bloch functions $\left\langle \mathbf{r}|1/2,s_{z}\right\rangle
$ (with $\mathbf{s}$ being the conduction band-edge angular momentum and $%
s_{z}=\pm 1/2$) times envelope functions. The later ones are obtained from
the effective mass Hamiltonian with a uniform electric field $|e|E_{QD}r\cos \theta$.
Here, $e$ is the electron charge $e$ and $E_{QD}$ is the electric
field intensity inside the nanocrystal.  At zero
electric field, the envelope functions take the form $R_{nl}(r)Y_{ll_{z}}(%
\theta ,\varphi )$. $R_{nl}(r)$ are the radial wave functions,\cite{e1} and $%
Y_{ll_{z}}(\theta ,\varphi )$ are the spherical harmonics.\cite{jackson}
These states are in the $l-s$ coupling scheme and have well defined values
of the squared orbital and spin angular momenta. Instead, we will use the $l-f$
coupling scheme, where the states have well defined total ($\mathbf{f}=%
\mathbf{l}+\mathbf{s}$) angular momentum projection $\hbar f_{z}$ and square
value $\hbar ^{2}f(f+1),$ that is
\begin{eqnarray}
\lefteqn{
\left\langle \mathbf{r}|nlff_{z}\right\rangle = } \nonumber \\
& &\sum_{l_{z},s_{z}}\left( l%
{\scriptstyle\frac{1}{2}}
l_{z}s_{z}|ff_{z}\right) R_{nl}(r)Y_{ll_{z}}(\theta ,\varphi
)\left\langle \mathbf{r}|1/2,s_{z}\right\rangle ,  \label{eq:elwf}
\end{eqnarray}
where $\left( l{\scriptstyle\frac{1}{2}}l_{z}s_{z}|ff_{z}\right) $ are the Clebsch-Gordan
coefficients, $f=l_{z}+s_{z}$, and $|l-1/2|\leq f\leq l+1/2$.\cite
{aggarwal74} If the hole states in the valence bands are described by an
spherical $4\times 4$ $\mathbf{k\cdot p}$ Hamiltonian, the $l-f$ coupling scheme for the
electron wave function is more convenient to build up the excitonic states
in a spherical QD. The hole and electron states present the same symmetry
properties, allowing to use all inherent properties of the Clebsch-Gordan
coefficients.\cite{brink}

Since the $s_{z}=\pm 1/2$ bands are uncoupled, the states described by Eq.~(%
\ref{eq:elwf}) have energies $E_{n,l}^{e}$ that are independent of the quantum
numbers $f,$ $f_{z}$. In the simulation of the real electronic states, the
confinement potential is chosen as a spherical box with an effective radius $%
R_{ef}$, which is greater than the structural nanocrystal radius $R$.
This effective radius is introduced in order to
take into account, approximately, the penetration of the electron wave
function in the surrounding medium. In our calculations, we determine $%
R_{ef} $ from the condition that the energy of the 1s state $E_{1,0}=\hbar
^{2}\pi ^{2}/2m_{e}R_{ef}^{2}$ be equal to the energy calculated for a
spherical well with depth $V_{e}=600$~meV.\cite{laheld}

The electron states under external electric fields are found by numerical
diagonalization of $H_{e}$ in the basis provided by (\ref{eq:elwf}). The matrix
elements of the Stark term are provided in the Appendix.

\begin{table}[!tbp]
\caption{Parameters used in the calculations.}
\label{tab3}%
\begin{ruledtabular}
\begin{tabular}{ll}
Parameter &  CdSe \\
  \hline
$E_{g}$ (eV) &  1.841\cite{landolt} \\
$m_{e}/m_{0}$ & 0.13\cite{laheld} \\
$\gamma _{1}$ & 1.66\cite{laheld} \\
$\gamma _{2}$ & 0.41\cite{laheld} \\
$2m_{0}P^{2}$ (eV)\footnotemark[1] & 20\cite{hermann} \\
$\epsilon $ &  9.53\cite{hraman} \\
$V_{e}$ (eV) & 0.6\cite{laheld} \\
$V_{h}$ (eV) & $\infty $\cite{laheld}
\end{tabular}
\end{ruledtabular}
\footnotetext[1]{$P=-i\left\langle S\left| \hat{p}_{x}\right|X\right\rangle /m_{0}$.}
\end{table}

For the hole states in the bottom of the valence band we use the well known
Baldereschi-Lipari Hamiltonian\cite{bald,baixia,efros92} in presence of a
constant electric field, that is
\begin{eqnarray}
H_{h}&=&\frac{\gamma _{1}}{2m_{0}}\left[ \mathbf{\hat{p}}^{2}-\frac{\mu }{9}%
\left( \mathbf{P}^{(2)}\cdot \mathbf{J}^{(2)}\right) \right] \nonumber \\
& &
+V_{h}(r)-|e|E_{QD}r\cos \theta .    \label{HamBL}
\end{eqnarray}
$V_{h}(r)$ being the confinement potential for the valence band, $\mathbf{P}%
^{(2)}$ and $\mathbf{J}^{(2)}$ are spherical tensors of rank 2 built from
linear and angular momentum operators, $\mu =2\gamma _{2}/\gamma _{1}$, and $%
\gamma _{2}$ and $\gamma _{1}$ are the Luttinger parameters of CdSe in the
spherical approximation $\gamma _{2}=\gamma _{3}$. The hole eigenfunctions
at zero electric field can be cast as
\begin{eqnarray}
\left\langle \mathbf{r}|NLFF_{z}\right\rangle
&=&\sum_{K=L,L+2}\sum_{L_{z},J_{z}}
\left( K{\scriptstyle\frac{3}{2}}L_{z}J_{z}|FF_{z}\right)   \nonumber \\
&&\times R_{N,K}^{(F)}(r)Y_{KL_{z}}(\theta ,\varphi )\left\langle \mathbf{r}%
|3/2,J_{z}\right\rangle ,  \label{eq:hlwf}
\end{eqnarray}
where $\left\langle \mathbf{r}|3/2,J_{z}\right\rangle $ are the hole Bloch
functions of the $\Gamma _{8}$ valence band with band-edge angular momentum $%
J=3/2$. The hole Bloch functions are related with the valence electron Bloch
functions $\left| \overline{J,J_{z}}\right\rangle $ by the rule $\left|
JJ_{z}\right\rangle =(-1)^{J-J_{z}}\left| \overline{J,-J_{z}}\right\rangle $%
\ (derived from the time-reversal operation). Our Bloch functions $\left|
3/2,J_{z}\right\rangle $ have the following phase convention: $\left| 3/2,\pm
3/2\right\rangle =\mp (i/\sqrt{2})(X\pm iY)\left| \pm \right\rangle $, and $%
\left| 3/2,\pm 1/2\right\rangle =(i/\sqrt{6})\left[ 2Z\left| \pm
\right\rangle \mp (X\pm iY)\left| \mp \right\rangle \right] $. The
phase factors of the above Bloch functions are implicit in the optical
dipole matrix elements.

For $F=1/2$, according to the rule of the addition of two angular
momenta, $|L-3/2|\leq F\leq L+3/2$,  the states defined in Eq. (\ref{eq:hlwf})
reduce to two
uncoupled states $L=1$ ($K=1$) and $L=2$ ($K=2$) which correspond to $%
P_{1/2} $ and $D_{1/2}$ states with radial wave functions $R_{N,L}^{(1/2)}$ (%
$L=1,2$). In this case the eigenfunctions $R_{N,L}^{(1/2)}$ fulfill two
independent radial effective mass equations with light hole effective
mass $m_{lh}=m_{0}/(\gamma_{1}+2\gamma _{2})$.

For $F\geq 3/2$, the radial wave functions $R_{N,K}^{(F)}(r)$ are solutions
of the coupled differential equations\cite{baixia,efros92}
\begin{widetext}
\begin{equation}
\label{eq:coupledR}
\left[
\begin{array}{cc}
-(1+C_1)\left( \frac{d^2}{dr^2}+\frac{2}{r}\frac{d}{dr}-
\frac{L(L+1)}{r^2}\right) +W(r,E)  &
C_2\left( \frac{d^2}{dr^2}+\frac{2L+5}{r}\frac{d}{dr}+
\frac{(L+1)(L+3)}{r^2}\right) \\
C_2\left( \frac{d^2}{dr^2}-\frac{2L+1}{r}\frac{d}{dr}+
\frac{L(L+2)}{r^2}\right) &
-(1+C_3)\left( \frac{d^2}{dr^2}+\frac{2}{r}\frac{d}{dr}-
\frac{(L+2)(L+3)}{r^2}\right) +W(r,E)
\end{array}
\right]
\left[ \begin{array}{c} R_{N,L}^{(F)}(r) \\R_{N,L+2}^{(F)}(r)\end{array}
\right]
=0,
\end{equation}
\end{widetext}
where $W(r,E)=2m_{0}\left[ V_{h}(r)-E\right] /\hbar ^{2}\gamma _{1}$.

The coefficients $C_{1}$, $C_{2}$ and $C_{3}$ are reported for several
states in Refs.~\onlinecite{bald} and \onlinecite{baixia}. Following the argument of
Baldereschi and Lipari, we have obtained the general expressions
\begin{subequations}
\begin{eqnarray}
C_{1}(L,F) &=&\mu \sqrt{5}(-1)^{3/2+L+F}\left\{
\begin{array}{ccc}
L & L & 2  \\
3/2 & 3/2 & F
\end{array}
\right\} \nonumber \\
&&\times \sqrt{\frac{2L(2L+1)(2L+2)}{(2L+3)(2L-1)}} ,  \\
C_{2}(L,F) &=&\mu \sqrt{30}(-1)^{3/2+L+F}\left\{
\begin{array}{ccc}
L+2 & L & 2 \\
3/2 & 3/2 & F
\end{array}
\right\} \nonumber \\
&&\times \sqrt{\frac{(L+1)(L+2)}{2L+3}}   , \\
C_{3}(L,F) &=&C_{1}(L+2,F)=-C_{1}(L,F).
\end{eqnarray}
\end{subequations}
The result $C_{3}=-C_{1}$ has been verified numerically. We also found numerically
that $C_{1}^2+C_{2}^2=\mu^2$ and $C_2/\mu>0$.

In the case of abrupt infinite confinement potential, the solutions of
equations (\ref{eq:coupledR}) are
\begin{subequations}
\begin{eqnarray}
\lefteqn{R_{N,L}^{(F)}(r)=A\left( j_{L}( kr/R)
- \frac{j_{L}(\sqrt{\beta}kr/R) j_{L}(k)}{j_{L}(\sqrt{%
\beta }k)}\right), }   \\
\lefteqn{
R_{N,L+2}^{(F)}(r)=-A\left(C_{1}/C_{2}+\sqrt{%
(C_1/C_2) ^{2}+1}\right)
}  \nonumber\\
&&\times \left( j_{L+2}(kr/R) -
\frac{j_{L+2}(\sqrt{\beta}kr/R) j_{L+2}(k)}{j_{L+2}(\sqrt{\beta }k)}\right),\quad
\end{eqnarray}
\end{subequations}
where $\beta =m_{lh}/m_{hh}$ ($m_{hh}=m_{0}/(\gamma _{1}-2\gamma _{2})$
is the heavy hole mass. The parameter $k$ fulfills the transcendental equation,
\begin{eqnarray}
\lefteqn{j_{L}(\sqrt{\beta }k)j_{L+2}(k)\left( C_{1}/C_{2}+\sqrt{%
(C_{1}/C_{2})^{2}+1}\right) =}  \nonumber \\
&&j_{L+2}(\sqrt{\beta }k)j_{L}(k)\left( C_{1}/C_{2}-
\sqrt{(C_{1}/C_{2})^{2}+1}\right) ,
\end{eqnarray}
and $A$ is a normalization constant, such that
\begin{equation}
\int \left( R_{N,L}^{(F)}(r)^{2}+R_{N,L+2}^{(F)}(r)^{2}\right) r^{2}dr=1.
\end{equation}
The hole energies are equal to $E_{h}=\hbar ^{2}k^{2}/2m_{hh}R^{2}$.

As in the case of the conduction band levels, the hole states under external
electric field are found by numerical diagonalization of the Hamiltonian $%
H_{h}$ in the basis provided by Eq.~(\ref{eq:hlwf}). The matrix element of the Stark
term are provided in the Appendix.

\begin{figure}[!b]
\includegraphics[angle=0, width=8.5cm]{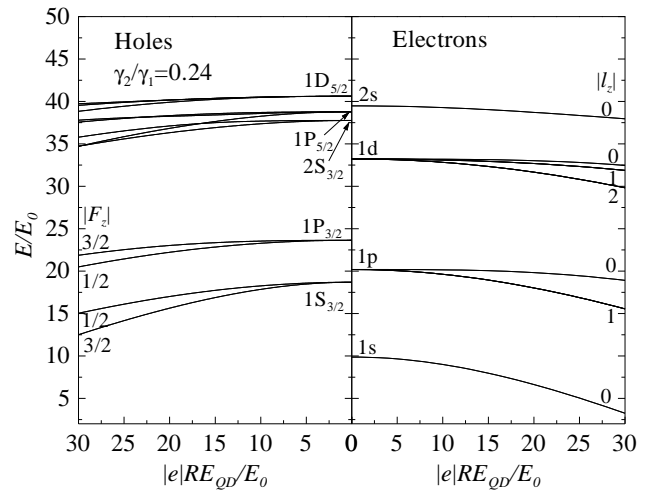}
\caption{Single particle energies vs. electric field intensity. Note that $%
E_{0}=\hbar^2/2m_i R^2$ depends upon the quasiparticle effective mass.}
\label{fig1}
\end{figure}

Figure~\ref{fig1} shows the energy levels of electrons and holes as functions
of the electric field intensity. The energies are in units of $E_{0}=\hbar
^{2}/2m_{i}R^{2}$ ($i=e,hh$), and the electric field intensity is in units
of $E_{0}/|e|R$. In this figure, $R$ for the electrons is the effective radius
above introduced.
The states at
zero field are indicated by the usual spectroscopy notation $NA$, with $%
A=S,P,D,...(s,p,d,...)$ for the $L=0,1,2,...(l=0,1,2...)$ hole (electron)
states. In the case of the hole energies, the quantum number $F$ is indicated
by a sub-index, that is, $NA_{F}.$ In the figure the values of $\left|
l_{z}\right| $ and $\left| F_{z}\right| $ for each states are indicated and,
as it can be seen, at non-zero electric field the $\pm \left| l_{z}\right| $
and the $\pm \left| F_{z}\right| $ degeneracy remains. For the electrons, due
to the absence of Bloch-envelope (spin-orbit-like) coupling, there is an
additional degeneracy. Thus, the electron energy levels  only depend upon the
modulus of the angular momentum projection $|l_{z}|$, being
degenerate in the spin projection. The results for the
electrons here shown reproduce those of Ref.~\onlinecite{casadostark}. The
light hole energies, which correspond to $F=1/2$, are higher and are not
included in Fig. 1.

The hole Hamiltonian  (\ref{HamBL}) does not include band warping terms that
arise from the cubic symmetry of the nanocrystal lattice.
The cubic corrections are proportional to the cubic coupling parameter \cite{bald74}
$\delta=(\gamma_3-\gamma_2)/\gamma_1$. The Luttinger parameter $\gamma_3$
has not been determined for CdSe, but it has been estimated as 0.53\cite{laheld} assuming
that the ratio $\gamma_3/\gamma_2$ is equal in CdTe and CdSe. Hence, a value
$\delta=0.066$ is obtained, which is one order of magnitude smaller than the spherical coupling
parameter $\mu$. Therefore, the cubic corrections can be included using perturbation theory.
For zinc-blende semiconductors, parity-breaking terms exist, in principle. However, this effect
seems to be smaller and the optical properties of acceptor levels have been explained
without considering them.\cite{bald74}
For zero electric field,
the hole states $P_{1/2}$ and $D_{1/2}$  belong to the irreducible representations
$\Gamma_6$ and $\Gamma_7$ of the point group $T_d$, respectively,
while  $S_{3/2}$ and  $P_{3/2}$ states belong to the irreducible representation
$\Gamma_8$. Hence, their energies do not split and the selection rules are not modified for these states.
However, the energy values shift in second order of $\delta$, unless coupled quasidegenerate
states are present.\cite{bald74} This is the case of the coupled states
$2S_{3/2}-1D_{5/2}$, where the splitting is first order in $\delta$.
For each $F>3/2$, the eigenstates of the cubic Hamiltonian are formed
of linear combinations of states
with different $F_z$. These linear combinations transform
according to the irreducible representations of $T_d$ that are compatible with
the representation $\mathcal{D}_F^{\pm}$ of the 3-dimensional rotation group $O(3)$.
Hence, the analyzed level splits proportionally to $\delta$
in several levels,
according to the compatibility tables of groups $T_d$ and $O(3)$.\cite{koster63}
 The optical selection rules are relaxed for these states and additional transitions should appear
 with low intensity.
 The effects of the cubic anisotropy  has been studied numerically
  for CdSe QDs
 using the Effective Bond Orbital Method (EBOM),\cite{laheld} which implicitly
 takes into account  the lack of inversion symmetry. It was shown that
 the spherical approximation is very good, specially for large nanocrystals.

\section{Exciton states}
 \label{sect:III}

 \begin{figure}[!tbh]
\includegraphics[angle=0, width=8.5cm]{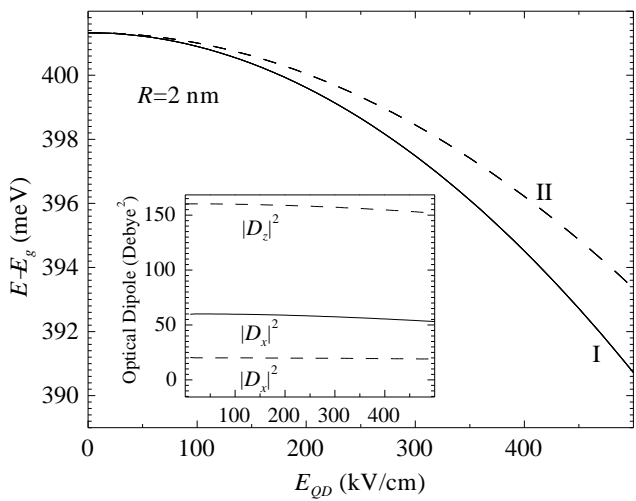}
\caption{Energy levels vs electric field of the lower excitonic states of a
CdSe nanocrystal 2 nm in radius. The inset shows the allowed optical dipole
matrix elements squared (see Sec. \protect\ref{sect:IV}). By symmetry the components $|D_y|^2$ and $|D_x|^2$
are equal.}
\label{fig3}
\end{figure}

The exciton Hamiltonian can be written as the sum of the electron and hole Hamiltonians
plus the screened Coulomb interaction
$V_{e-h}=-e^{2}/\epsilon \left| \mathbf{r}_{e}-\mathbf{r}_{h}\right|$,
 $\epsilon $ being the dielectric constant. Excitonic states, can be
obtained using an expansion in a basis of electron-hole pair wave functions
with well defined total angular momentum square $\hbar ^{2}M(M+1)$ and
projection $\hbar M_z$
\begin{equation}
\left| \Psi \right\rangle =\sum_{\alpha =\{n,N,l,L,f,F,M,M_{z}\}}C_{\alpha
}\left| nNlLfF;MM_{z}\right\rangle ,  \label{eq:exwf}
\end{equation}
where
\begin{eqnarray}
\lefteqn{\left| nNlLfF;MM_{z}\right\rangle }  \nonumber \\
&=&\sum_{f_{z},F_{z}}\left( fFf_{z}F_{z}|MM_{z}\right) \left|
nlff_{z}\right\rangle \otimes \left| NLFF_{z}\right\rangle .
\label{eq:ehpwf}
\end{eqnarray}
It is important to remark that in Eqs.~(\ref{eq:exwf}) and (\ref{eq:ehpwf})
it is implicit the condition for the addition of two angular momenta for
electrons and holes $|f-1/2|\leq l\leq f+1/2$ and $|F-3/2|\leq L\leq F+3/2,$
respectively$.$  The matrix elements of the
Coulomb interaction in the basis (\ref{eq:exwf}) are
reported in Ref.~\onlinecite{articlehr}. The matrix elements of the Stark
term are provided in the Appendix.

As the electric field is chosen along the Z-axis, the exciton angular
momentum Z-projection is a constant of motion and the Hamiltonian can be
diagonalized independently in different $M_{z}$-subspaces. We have built the
$M_{z}$-subspaces using as basis all the possible electron-holes states
(\ref{eq:ehpwf}) that fulfill the condition $2m_{i}R^{2}E/\hbar ^{2}\leq 90$
($i=e,hh$). With these criteria the dimensions of the diagonalized
matrices are 532, 502, and 415 for $|M_{z}|=0$, $1$, and $2$, respectively.

Figure~\ref{fig3} shows the structure of the lower exciton energy level as a
function of the applied electric field $E_{QD}$ in a CdSe nanocrystal 2~nm
in radius. The lower exciton at zero electric field is 8-fold degenerate: 3
states with $M=1$ and 5 states with $M=2$. All these states are originated
from the 1s-1S$_{3/2}$ electron-hole pairs. The electric field splits this
level in two quartets: the lower one (I) belongs to states with $M_{z}=\pm
1,\pm 2$ while the higher level (II) in Fig. \ref{fig3} corresponds to exciton
wave function with $M_{z}=0,0,\pm 1$. The inset in Fig.~\ref{fig3} displays
the matrix elements of the dipole operator, which determine the optical properties
and will be discussed in the next section.
Table~\ref{tab1} illustrates the
fraction contribution, $\left| C_{M}(M_{z})\right| ^{2},$ of the dominant $M$%
-components in the expansion (\ref{eq:exwf}) to the excitonic levels I and
II at $E_{QD}=500$ kV/cm. It is worth to note that the states with
$M_{z}=\pm 1$ are derived from the $M=1$ and $2$  zero-field exciton wave
functions, which are coupled by the electric field through remote states. The
following group of exciton levels arises from the 1s-1P$_{3/2}$
electron-hole pairs, with a splitting pattern similar to that of the lower
level. In general, all the $M_{z}=0$ levels are doubly degenerate. It is
important to remark that the above description is valid for larger nanocrystals
in the strong confinement regime.

\begin{figure}[!tbh]
\includegraphics[angle=0, width=8.5cm]{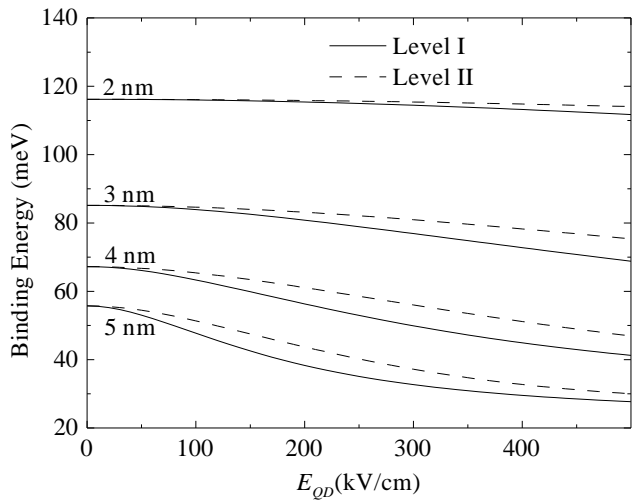}
\caption{Exciton binding energy versus electric field for the levels I and
II shown in Fig. \ref{fig3} for four QD radius (2,3,4, and 5 nm).}
\label{fig4}
\end{figure}

The dependence of the binding energy upon $E_{QD}$, corresponding to the I and
II exciton levels are shown in Fig. \ref{fig4} for several nanocrystal
radii.

\begin{figure}[!tbh]
\includegraphics[angle=0, width=8.5cm]{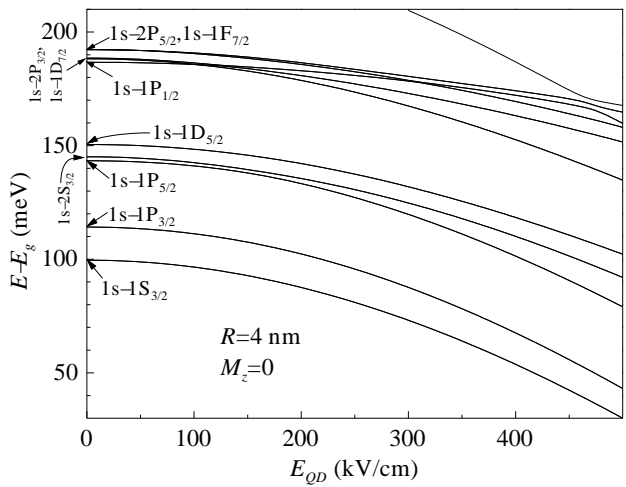}
\caption{Exciton energies with quantum number $M_{z}=0$ vs electric field
intensity for a CdSe nanocrystal 4~nm in radius.}
\label{fig6}
\end{figure}

The energy levels with $M_{z}=0$, up to 100 meV above the lowest exciton, are
plotted in Fig.~\ref{fig6}. The states are labeled by the pure electron-hole
pair contribution at zero field. Notably, no anti-crossing behavior is
observed in levels 1s-1S$_{3/2}$ and 1s-1P$_{3/2}$, and there is a weak
anti-crossing  between 1s-1D$_{5/2}$ and 1s-2S$_{3/2}$ in the region 100-150 kV/cm.
The anticrossings should not be altered by the effect of the cubic terms in the hole
Hamiltonian, although the absolute values will change.

\section{Field induced optical properties}
\label{sect:IV}
In the optical experiments the intensity of the absorbed or emitted
light is proportional to the exciton oscillator strength (squared absolute values of the
dipole matrix element between the ground and excited states of the quantum
dot). The field will modify the electron and hole wave
functions which  determine the exciton oscillator strength.
As the field
delocalizes the electron and holes in opposite directions, in principle the
allowed excitonic transitions should decrease as the applied electric field
increases. This is related to the quenching of the luminescence spectra.%
\cite{Kash,Miller83} Moreover, the field breaks the inner spherical symmetry of
the dot and forbidden excitonic transitions should appear in the optical
spectra for large electric fields.

\begin{table}[!tb]
\caption{Fraction of the dominant $M$-component $|C_M(M_z)|^2$ to the
expansion (\ref{eq:exwf}) for the lower excitonic states I and II shown in
Fig.~\ref{fig3} at the electric field intensity $E_{QD}=500$~kV/cm. }
\label{tab1}%
\begin{ruledtabular}
\begin{tabular}{ldd|ddd}
Level &\multicolumn{2}{c|}{I}&\multicolumn{3}{c}{II}\\
\hline
$E-E_g$ &\multicolumn{2}{c|}{390.7~meV}&\multicolumn{3}{c}{393.4~meV}\\
\hline
 & \multicolumn{2}{c|}{$|C_M(M_z)|^2$-fraction} & \multicolumn{3}{c}{$|C_M(M_z)|^2$-fraction} \\
\hline
 $M\backslash M_z$ & (\pm 2) & (\pm 1) & 0     & 0 & (\pm 1) \\
\hline
$0$   &0      &0      &0      &0.01   &0      \\
$1$   &0      &0.74   &0.02   &0.96   &0.25  \\
$2$   &0.99   &0.25   &0.96   &0.03   &0.74\\
$3$   &0.01   &0      &0.02   &0      &0.01
\end{tabular}
\end{ruledtabular}
\end{table}

The optical matrix element of the dipole operator for the exciton
(\ref{eq:exwf}) is given by
\begin{equation}
\mathbf{D}_{\Psi ,G}=\sum_{\alpha=\{n,N,l,L,f,F;M,M_z \} }
C_{\alpha }(\Psi )^{\ast }\mathbf{D}%
_{\alpha ,G},  \label{eq:dip}
\end{equation}
where $G$ represents the nanocrystal ground state and
$\mathbf{D}_{\alpha ,G}=i|e|\hbar\langle \alpha |%
\hat{\mathbf{p}}|G\rangle/m_{0}(E_{\alpha}-E_{G})$
 are the
dipole matrix elements for the uncorrelated electron-hole pairs
$\alpha $ (Eq.~\ref{eq:ehpwf}),
which are given in Ref.~\onlinecite{articlehr}.
These dipole matrix element are different from zero only for $M=1$ and
are proportional to
$\mathbf{\hat{e}}_{M_{z}}^{*}$, where
$\mathbf{\hat{e}}_{M_{z}=0}=\mathbf{\hat{e}}_z$ and
$\mathbf{\hat{e}}_{M_{z}=\pm 1}=\mp\left(\mathbf{\hat{e}}_x\pm%
i\mathbf{\hat{e}}_y\right)/\sqrt{2}$
are the unit vectors in the spherical
representation. This  accounts for the optical selection rules in the
dipole approximation: the only optically active electron-hole pairs have
$M_{z}=0,\pm 1$, and $M=1$. According to the results of Table~\ref{tab3}, the
exciton level I in Fig.~\ref{fig3} is optically active for light polarized
(with electric polarization vector $\mathbf{e}_l$)
along $\mathbf{\hat{e}}_x$ or equivalently
$\mathbf{e}_{l}\parallel \mathbf{\hat{e}}_{M_{z}=\pm 1}$,
while the state II is allowed for
$\mathbf{e}_{l}\parallel \mathbf{\hat{e}}_{z}\parallel \mathbf{\hat{e}}_{M_{z}=0}$
or  $\mathbf{e}_{l}\parallel \mathbf{e}_{x}\parallel \mathbf{\hat{e}}_{M_{z}=\pm 1}$.
The corresponding
squares of the dipole matrix elements $|D_{z}=\mathbf{\hat{e}}_{z}\cdot\mathbf{D}%
_{\Psi ,G}|^{2}$ and $|D_{x}=\mathbf{\hat{e}}_{x}\cdot\mathbf{D}_{\Psi ,G}|^{2}$
for the states I and II are shown in the inset of Fig.~\ref{fig3}.
Note that at zero electric field the sum of $|D_{x}|^{2}$ over all $%
M_{z}=\pm 1$ states equals to $|D_{z}|^{2}$, as required by the spherical
symmetry of our Hamiltonian.

\begin{figure}[!tbh]
\includegraphics[angle=0, width=8.5cm]{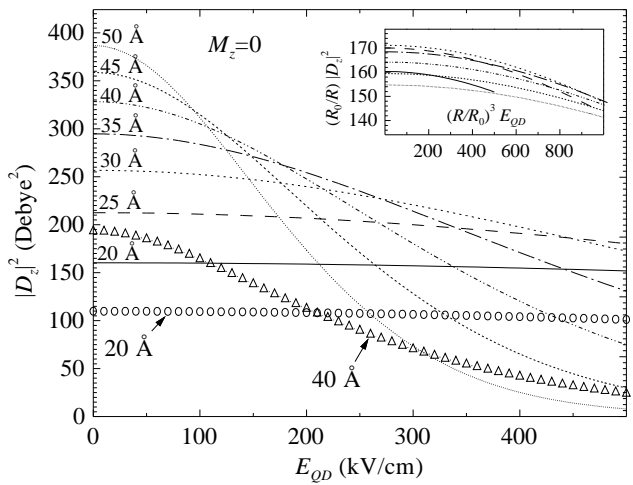}
\caption{Allowed dipole matrix elements, $|D_{z}|^{2}$, of the lower
excitonic states II as function of the applied electric field for CdSe
nanocrystals of several radii.}
\label{fig5}
\end{figure}

\begin{figure}[!tbh]
\includegraphics[angle=0, width=8.5cm]{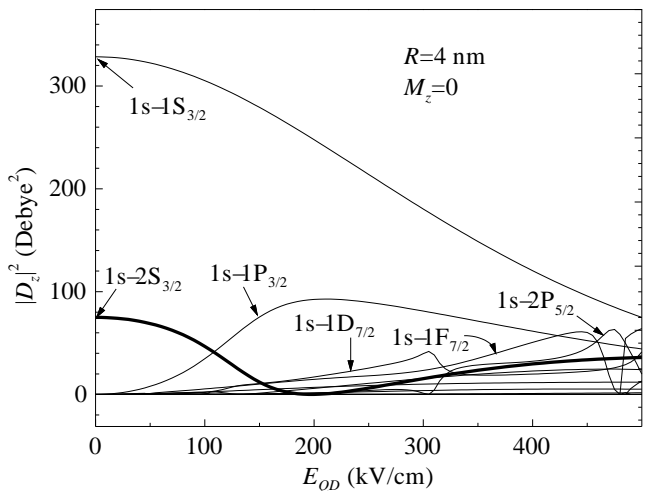}
\caption{Dipole matrix element $|D_{z}|^{2}$ for a CdSe nanocrystal 4~nm in
radius as function of $E_{QD}$ for different excitonic states with angular
momentum projection $M_{z}=0$.}
\label{fig7}
\end{figure}

Figure~\ref{fig5} shows the dependence of $|D_{z}|^{2}$ upon the electric
field intensity for the lower excitonic states with quantum number $M_{z}=0$
for CdSe nanocrystals of several radii. In the inset we present the same
results shown in the figure but rescaled according to the laws $%
(R_{0}/R)|D_{z}|^{2}$ and $(R/R_{0})^{3}E_{QD},$ where $R_{0}=2$~nm.
It can be noticed that $|D_{z}|^{2}$ scales almost as $R$, while the
electric field scales as $R^{-3}$. The linear dependence of
$|D_{z}|^{2}$ upon $R$ is a consequence of the Coulomb interaction, i.e.
 an excitonic effect.
For sake of
comparison in the Fig.~\ref{fig5} the free electron-hole calculations of
$|D_{z}|^{2}$ are shown by the triangles and
circles for dot radii equal to 2~nm and
4~nm, respectively. As can be seen in the figure, the exciton effects are
large, even for small radius of 2~nm. This means that the usual
strong confinement approximation, where the Coulomb interaction is
considered as a small perturbation, breaks down for small CdSe
nanocrystals. This effect can be explained due to the finite confinement
barrier of the electron allowing the penetration of the exciton wave
function in the surrounding medium. Also, Fig.~\ref{fig5} indicates that for
small radius the optical dipole is not quenched, while for large radius the
reduction in photoluminescence intensity should be significant. For example,
for a field of 150~kV/cm the square dipole matrix element decreases
approximately by 66\% for a QD 50~\AA\ in radius, while for a
QD of 20 \AA , $|D_{z}|^{2}$ is almost a constant for the range
of the experimental values of the electric field that can be considered. The
explanation of the above considered features lies in the interplay between
the confined and electric energies, i.e. the kinetic energy and the
external potential energy depend as $R^{-2}$ and $R$, respectively. In Fig.~%
\ref{fig7}, the behavior of the oscillator strength $|D_{z}|^{2}$ as a
function of $E_{QD}$ is shown for several excitonic states with angular momentum
projection $M_{z}=0$. At $E_{QD}=0$ the exciton oscillator strength is diagonal
and only transitions between electrons and holes with $S$- symmetry are
allowed. The field breaks this selection rule and transitions with different
symmetry are allowed, i.e. the electric field couples states with $L$,
$l\neq 0$ and the oscillator strength becomes different from zero for
$s-P$, $s-D$, ... electron-hole pair transitions. The mixing effect, due
to $\mathbf{E}_{QD},$ reduces the overlapping between electron and hole
states with $S$- symmetry, allowing other electron-hole transitions. In
Fig.~\ref{fig7}, the transition 1s-1S$_{3/2}$ is the strongest one over the
full range of electric field intensity. For higher transitions, there are kinks
at 300 and 480~kV/cm,
which are due to avoided crossings between higher levels, e.g.,
between 1s-1D$_{7/2}$ and 1s-1F$_{5/2}$ at 300~kV/cm (see Fig.~\ref{fig6}).
Another interesting feature is the
disappearance and the appearance of several transitions as the field is
tuned. For example, the dipole of the transition $1s-2S_{3/2}$, shown as thick line
in Fig.~\ref{fig7}, is strong at
$E_{QD}=0$ and disappears at $E_{QD}\sim 180-220$ kV/cm.
 The opposite can
be argued for the $1s-1D_{7/2}$ and $1s-1F_{7/2}$ transitions, for which the
dipole matrix elements reach a maximum and at higher field intensities they
have practical zero oscillator strength. Furthermore, the transition $1s-2S_{3/2}$ reappears
at higher $E_{QD}$.

\section{Discussion}

Our calculation reproduces the magnitude of the Stark shift, which can be as
large as 80~meV for internal field intensity of 500~kV/cm in nanocrystals 4~%
nm in radius. This value agrees well in magnitude with the experimental
observations of Ref.~\onlinecite{science97} if one considers that the maximum of
the optical gap indicates the zero of the internal electric field under an
applied external potential. The presence of  an internal electric field and
other features of single dot luminescence are currently attributed to
trapped charges near the surface of the nanocrystal.%
\cite{science97,wang01,eychm00,title97,shim01}
 Large dipole moments for the ground
state and the LUMO in intrinsic CdSe nanocrystals were predicted by a set of
pseudopotential calculations,\cite{rabani99} which can also account for the
linear Stark shift in single dots. This intrinsic dipole moment is
associated with the lack of inversion symmetry in wurtzite lattice and
depends upon dot structural details and  the dielectric response of the
surrounding medium. The same calculation also reports null dipole moments
for CdSe dots with zinc-blende structure, thus supporting our theoretical
model.

Our results are qualitatively similar to those of Fu\cite{fu02} for InP QD, with the
exception of non zero dipole moment at zero field. Due to the inversion symmetry of our hole
Hamiltonian (\ref{HamBL}), the hole eigenstates have definite parity
and zero dipole moment. A
finite dipole moment, caused by the lack of inversion symmetry in the zinc-blende structure,
can be obtained if a linear term in $\mathbf{k}$ is
included in the $\mathbf{k}\cdot \mathbf{p}$ Hamiltonian (\ref{HamBL}). This
effect is tiny in bulk semiconductors and is usually neglected. In
nanocrystals the dipole may arise from mixing of the hole states $1S_{3/2}$
and $1P_{3/2}$. The amount of mixing depends upon the ratio of the $\mathbf{k}$
linear term matrix element ($\propto 1/R$) to the energy separation of the
level ($\propto 1/R^{2}$). Hence, the dipole should be small for small
nanocrystals with zinc-blende lattice. For large nanocrystals the bulk regime
is approached and the linear terms are again negligible. Although it is not
possible to predict with certitude the dipole behavior for intermediate dot
sizes, its absolute value should not grow. This is consistent with the
calculations of Refs.~\onlinecite{rabani99} and \onlinecite{fu02}.

An  anomalous field dependence of
emission lines of self-assembled quantum dots (SAQD)  has been observed through
micro-photoluminescence measurements,\cite{raydmond} where certain transitions
lines of  the luminescence spectrum disappear and reappear as the electric field
is tuned. This feature could not be accounted for using a 1D model of the quantum
confinement.\cite{raydmond} The dipole matrix element of the state 1s-2S$_{3/2}$
(Fig.~\ref{fig7}) displays that behavior, which
could be a property of the 3D confinement. However, a detailed calculation with
the SAQD symmetry is needed to test this hypothesis.

\begin{figure}[!tbh]
\includegraphics[angle=0, width=8.5cm]{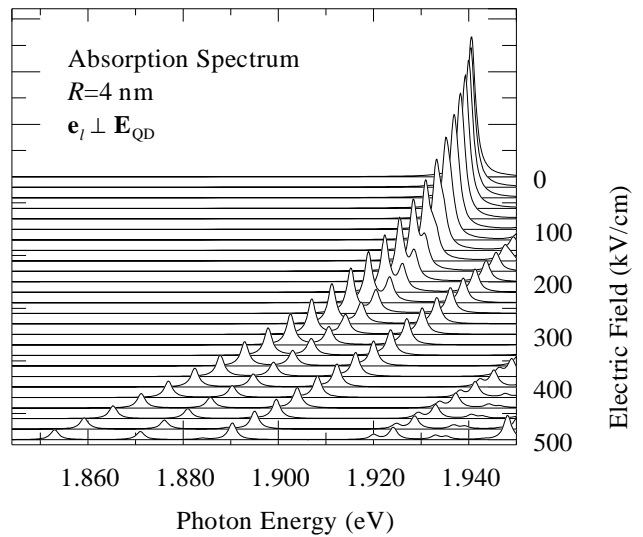}
\caption{Absorption spectra for a CdSe nanocrystal at different electric
field intensities. The light wave vector and its polarization
$\mathbf{e}_l$ are perpendicular to the electric field.}
\label{fig8}
\end{figure}

Let us consider two simple configurations for photoluminescence
experiments. First, the emitted light, with wave vector $\bm{\kappa}$, is recorded
along the $\mathbf{E}_{QD}$ direction and its polarization vector $\mathbf{e}_{l}$
is parallel to $\mathbf{\hat{e}}_{x},$ (that is,
$\mathbf{e}_{l}\parallel \mathbf{\hat{e}}_{x}\perp\mathbf{E}_{QD}$).
In this case the dipole element
$D_{x}=\mathbf{\hat{e}}_{x}\cdot\mathbf{D}_{\Psi,G}$ is not zero
and, according to Table \ref{tab3}, the states with
$M_{z}=\pm 1$ of the lowest excitonic level are optically active.
Photoluminescence spectra should provide the Stark splitting presented in
Fig.~\ref{fig3}.
Second, the emitted light is observed perpendicular to the field
($\bm{\kappa}\perp\mathbf{E}_{QD}$). Here, we have two choices for the light
polarization, (i) $\mathbf{e}_{l}\parallel \mathbf{E}_{QD}$ or (ii)
$\mathbf{e}_{l}\perp \mathbf{E}_{QD}$.  In the case (i),
$\mathbf{e}_{l}\parallel \mathbf{\hat{e}}_{z}\parallel \mathbf{\hat{e}}_{M_{z}=0}$
and the emitted light corresponds to the excitonic state II and the
photoluminescence spectrum presents only one peak.
 In the configuration (ii), both excitonic states I and II are
activated and the Stark splitting of Fig. \ref{fig3} appears.
The case (ii) is illustrated in Fig.~\ref{fig8} for the absorption spectra
of a single QD at different electric fields. In the figure, the quenching of  the absorption
lines and the split of the 1s-1S$_{3/2}$ exciton peak as the electric field
increases are clearly observed.

The above picture is slightly modified by the cubic anisotropy. At zero electric field,
the 8-fold degenerate 1s-1S$_{3/2}$ excitons belong to the representations
$ \mathcal{D}^{-}_{1} $ ($M=1$) and   $ \mathcal{D}^{-}_{2} $ ($M=2$)
of group $O(3)$. It is remarkable that the Coulomb
interaction does not remove the degeneracy  of $ \mathcal{D}^{-}_{1} $  and
$ \mathcal{D}^{-}_{2}$, and neither does it for higher levels.
 According to the compatibility tables for groups $O(3)$ and
 $T_d$,\cite{koster63}  $\mathcal{D}^{-}_{1}=\Gamma_5$ and
 $\mathcal{D}^{-}_{2}=\Gamma_3 + \Gamma_4$.
 Hence, the $M=1$ triplet is not splitted.  As only
 $\Gamma_5$ is dipole allowed, the selection rule $M=1$ remains valid for the
 ground exciton state. Only exciton states  $n$s-$N$S$_{3/2}$  belong to
  $ \mathcal{D}^{-}_{1} $ and are dipole allowed in the spherical approximation.
  The rest of the states, except  $n$s-$N$P$_{1/2}$, split in different levels that include
    $\Gamma_5$ and should produce weak lines in the optical spectra.
   For non-zero electric field, the $O(3)$ symmetry is reduced to $C_{\infty v}$, and the
 irreducible representations correspond to the different values of $|M_z|$. The effect of the
 cubic terms depend upon the orientation of the crystal axes relative to the electric field. If
 $\mathbf{E}_{QD}$ is parallel to a 3-order axis, the symmetry is reduced to $C_{3v}$.
 Hence, the excitons with $M_z=0$ and $M_z=\pm 1$ belong to the  $\Gamma_1$ and $\Gamma_3$
 irreducible representations of $C_{3v}$, respectively, and are not modified by the symmetry reduction.
 Excitons with higher $|M_z|$ can split and give rise
 to weak optical transitions. If the electric field is not oriented along a 3-order axis, all the degeneracies
 are removed  and additional transitions should appear.
However, as the cubic anisotropy
 is induced fundamentally through the hole states, the splittings and intensities of the
 extra lines should be extremely  small.

The fine structure that we have revealed can be modified by other effects present in
real nanocrystals, such as shape asymmetry, crystal field
(in wurtzite nanocrystals), and the electron-hole exchange interaction.\cite{rosen}
 These effects lead to a splitting pattern
of the 8-fold lowest exciton into 5-levels, which is combined with the electric field-induced splitting.
 Non-adiabatic phonon-induced effects\cite{fomin}
and dielectric mismatch-induced modifications of the Coulomb interaction\cite{banyai92}
contribute also to the magnitude of the splittings. For CdSe QDs with wurtzite structure
the relative orientation of the crystallographic axes and the external electric field must
have observable signatures in the optical spectra. These effects may be stronger than those
produced by the cubic anisotropy. Since our approach ignores these effects
our discussion is approximate.

In summary, we have studied the influence of the electric field on the electron and hole
single particle states in CdSe nanocrystals, as well as on the exciton states and
the optical properties. We have described the electric-field-induced
quenching of the absorption and luminescence spectra
and the importance of the exciton
effects. We have found  that the Coulomb interaction has a large
influence on the strength of the optical
transitions, even for small quantum dots. This fact is related to the penetration of electron
wave function in the embedding medium, that partially breaks the strong confinement regime.
Moreover, we have shown that for small QDs the dependence of the exciton binding energy
upon the applied electric field is strongly reduced.
For zero electric field we have reported very general expressions for the solutions
of the 4$\times 4$ hole Hamiltonian in spherical QDs.

\begin{acknowledgments}
This work was partially supported by Alma Mater project 26-2000 of Havana
University.
\end{acknowledgments}

\appendix
\section{Matrix elements of the Stark term}

The Stark term in the electron effective mass Hamiltonian is a component of
an irreducible spherical tensor of rank 1.
The theorem of Wigner-Eckart and
the reduction formulas for compound systems
detailed in Ref.~\onlinecite{brink} allows us to write the matrix elements as
\begin{eqnarray}
\lefteqn{\left\langle n^{\prime} l^{\prime} f^{\prime} f_{z}^{\prime} \left|
|e|E_{QD}r\cos \theta \right| nlff_{z}\right\rangle }  \nonumber \\
&=&(-1)^{f^{\prime }-f_{z}^{\prime }+3/2+f}|e|E_{QD}\sqrt{(2f^{\prime
}+1)(2f+1)}  \nonumber \\
&&\times \left(
\begin{array}{ccc}
f^{\prime } & 1 & f \\
-f_{z}^{\prime } & 0 & f_{z}
\end{array}
\right) \left\{
\begin{array}{ccc}
f^{\prime } & f & 1 \\
l & l^{\prime } & 1/2
\end{array}
\right\} \left(
\begin{array}{ccc}
l^{\prime } & 1 & l \\
0 & 0 & 0
\end{array}
\right)  \nonumber \\
&&\times \sqrt{(2l^{\prime }+1)(2l+1)}\int R_{n^{\prime }l^{\prime
}}(r)R_{nl}(r)r^{3}dr,
\end{eqnarray}
where the terms inside $\{\}$ and $()$ are the Wigner's 6-j and 3-j symbols,
respectively.

For the hole states we have found the expression
\begin{eqnarray}
\lefteqn{\left\langle N^{\prime }L^{\prime }F^{\prime }F_{z}^{\prime }\left|
-|e|E_{QD}r\cos \theta \right| NLFF_{z}\right\rangle }  \nonumber \\
&=&(-1)^{F^{\prime }-F_{z}^{\prime }+3/2+F}|e|E_{QD}\sqrt{(2F^{\prime }+1)(2F+1)}
\nonumber \\
&&\times \left(
\begin{array}{ccc}
F^{\prime } & 1 & F \\
-F_{z}^{\prime } & 0 & F_{z}
\end{array}
\right) \sum_{K,K^{\prime }}\left\{
\begin{array}{ccc}
F^{\prime } & F & 1 \\
K & K^{\prime } & 3/2
\end{array}
\right\} \left(
\begin{array}{ccc}
K^{\prime } & 1 & K \\
0 & 0 & 0
\end{array}
\right)  \nonumber \\
&&\times \sqrt{(2K^{\prime }+1)(2K+1)}\int R_{N^{\prime },K^{\prime
}}^{(F^{\prime })}(r)R_{N,K}^{(F)}(r)r^{3}dr.
\end{eqnarray}

\begin{widetext}
For the exciton states we found the expression
\begin{eqnarray}
\lefteqn{\left\langle n'N'l'L'f'F';M'M_z'\left|
|e|E_{QD}(r_e\cos\theta_e-r_h\cos\theta_h)\right|nNlLfF;MM_z\right\rangle
}\nonumber \\
&=& |e|E_{QD} (-1)^{M'-M_z'}\delta_{M_z,M_z'}
\left(\begin{array}{ccc} M'&1&M\\-M_z'&0&M_z \end{array}\right)
\sqrt{(2M+1)(2M'+1)}
\Bigg\{
\delta_{N,N'}\delta_{L,L'}\delta_{F,F'}(-1)^{F+M+f+f'+1/2} \nonumber \\
& & \times \sqrt{(2f'+1)(2f+1)(2l'+1)(2l+1)}
\left\{\begin{array}{ccc} f'&f&1\\l&l'&1/2 \end{array} \right\}
\left\{\begin{array}{ccc} M'&M&1\\f&f'&F \end{array} \right\}
\left(\begin{array}{ccc} l'&1&l\\0&0&0 \end{array} \right)
\int R_{n'l'}(r_e) R_{nl}(r_e) r_e^3 dr_e \nonumber \\
& & -\delta_{n,n'}\delta_{l,l'}\delta_{f,f'}(-1)^{M'+f+3/2+2F}\sqrt{(2F'+1)(2F+1)}
\sum_{K,K'}\sqrt{(2K'+1)(2K+1)} \nonumber \\
& & \times
\left\{\begin{array}{ccc} F'&F&1\\K&K'&3/2 \end{array} \right\}
\left\{\begin{array}{ccc} M'&M&1\\F&F'&f \end{array} \right\}
\left(\begin{array}{ccc} K'&1&K\\0&0&0 \end{array} \right)
\int R_{N',K'}^{(F')}(r_h) R_{N,K}^{(F)}(r_h) r_h^3 dr_h
\Bigg\}.
\end{eqnarray}
\end{widetext}


\begin{thebibliography}{43}
\expandafter\ifx\csname natexlab\endcsname\relax\def\natexlab#1{#1}\fi
\expandafter\ifx\csname bibnamefont\endcsname\relax
  \def\bibnamefont#1{#1}\fi
\expandafter\ifx\csname bibfnamefont\endcsname\relax
  \def\bibfnamefont#1{#1}\fi
\expandafter\ifx\csname citenamefont\endcsname\relax
  \def\citenamefont#1{#1}\fi
\expandafter\ifx\csname url\endcsname\relax
  \def\url#1{\texttt{#1}}\fi
\expandafter\ifx\csname urlprefix\endcsname\relax\def\urlprefix{URL }\fi
\providecommand{\bibinfo}[2]{#2}
\providecommand{\eprint}[2][]{\url{#2}}

\bibitem[{\citenamefont{Mendez et~al.}(1982)\citenamefont{Mendez, Bastard,
  Chang, Esaki, Morkoc, and Fischer}}]{mendez}
\bibinfo{author}{\bibfnamefont{E.~E.} \bibnamefont{Mendez}},
  \bibinfo{author}{\bibfnamefont{G.}~\bibnamefont{Bastard}},
  \bibinfo{author}{\bibfnamefont{L.~L.} \bibnamefont{Chang}},
  \bibinfo{author}{\bibfnamefont{L.}~\bibnamefont{Esaki}},
  \bibinfo{author}{\bibfnamefont{H.}~\bibnamefont{Morkoc}}, \bibnamefont{and}
  \bibinfo{author}{\bibfnamefont{R.}~\bibnamefont{Fischer}},
  \bibinfo{journal}{Phys. Rev. B} \textbf{\bibinfo{volume}{26}},
  \bibinfo{pages}{7101} (\bibinfo{year}{1982}).

\bibitem[{\citenamefont{Miller et~al.}(1984)\citenamefont{Miller, Chemla,
  Damen, Gossard, Wiegmann, Wood, and Burrus}}]{miller}
\bibinfo{author}{\bibfnamefont{D.~A.~B.} \bibnamefont{Miller}},
  \bibinfo{author}{\bibfnamefont{D.~S.} \bibnamefont{Chemla}},
  \bibinfo{author}{\bibfnamefont{T.~C.} \bibnamefont{Damen}},
  \bibinfo{author}{\bibfnamefont{A.~C.} \bibnamefont{Gossard}},
  \bibinfo{author}{\bibfnamefont{W.}~\bibnamefont{Wiegmann}},
  \bibinfo{author}{\bibfnamefont{T.~H.} \bibnamefont{Wood}}, \bibnamefont{and}
  \bibinfo{author}{\bibfnamefont{C.~A.} \bibnamefont{Burrus}},
  \bibinfo{journal}{Phys. Rev. Lett.} \textbf{\bibinfo{volume}{53}},
  \bibinfo{pages}{2173} (\bibinfo{year}{1984}).

\bibitem[{\citenamefont{Kapteyn et~al.}(1999)\citenamefont{Kapteyn,
  Heinrichsdorff, Stier, Heitz, Grundmann, Zakharov, Bimberg, and
  Werner}}]{kapteyn}
\bibinfo{author}{\bibfnamefont{C.~M.~A.} \bibnamefont{Kapteyn}},
  \bibinfo{author}{\bibfnamefont{F.}~\bibnamefont{Heinrichsdorff}},
  \bibinfo{author}{\bibfnamefont{O.}~\bibnamefont{Stier}},
  \bibinfo{author}{\bibfnamefont{R.}~\bibnamefont{Heitz}},
  \bibinfo{author}{\bibfnamefont{M.}~\bibnamefont{Grundmann}},
  \bibinfo{author}{\bibfnamefont{N.~D.} \bibnamefont{Zakharov}},
  \bibinfo{author}{\bibfnamefont{D.}~\bibnamefont{Bimberg}}, \bibnamefont{and}
  \bibinfo{author}{\bibfnamefont{P.}~\bibnamefont{Werner}},
  \bibinfo{journal}{Phys. Rev. B} \textbf{\bibinfo{volume}{60}},
  \bibinfo{pages}{14265} (\bibinfo{year}{1999}).

\bibitem[{\citenamefont{Miller et~al.}(1988)\citenamefont{Miller, Chemla, and
  Schmitt-Rink}}]{miller88}
\bibinfo{author}{\bibfnamefont{D.~A.~B.} \bibnamefont{Miller}},
  \bibinfo{author}{\bibfnamefont{D.~S.} \bibnamefont{Chemla}},
  \bibnamefont{and}
  \bibinfo{author}{\bibfnamefont{S.}~\bibnamefont{Schmitt-Rink}},
  \bibinfo{journal}{Appl. Phys. Lett.} \textbf{\bibinfo{volume}{52}},
  \bibinfo{pages}{2154} (\bibinfo{year}{1988}).

\bibitem[{\citenamefont{Raymond et~al.}(1998)\citenamefont{Raymond, Reynolds,
  Merz, Fafard, Feng, and Charbonneau}}]{raydmond}
\bibinfo{author}{\bibfnamefont{S.}~\bibnamefont{Raymond}},
  \bibinfo{author}{\bibfnamefont{J.~P.} \bibnamefont{Reynolds}},
  \bibinfo{author}{\bibfnamefont{J.~L.} \bibnamefont{Merz}},
  \bibinfo{author}{\bibfnamefont{S.}~\bibnamefont{Fafard}},
  \bibinfo{author}{\bibfnamefont{Y.}~\bibnamefont{Feng}}, \bibnamefont{and}
  \bibinfo{author}{\bibfnamefont{S.}~\bibnamefont{Charbonneau}},
  \bibinfo{journal}{Phys. Rev. B} \textbf{\bibinfo{volume}{58}},
  \bibinfo{pages}{R13415} (\bibinfo{year}{1998}).

\bibitem[{\citenamefont{Fry et~al.}(2000)\citenamefont{Fry, Itskevich, Mowbray,
  Skolnick, Finley, Barker, O{'}Reilly, Wilson, Larkin, Maksym et~al.}}]{Fry}
\bibinfo{author}{\bibfnamefont{P.~W.} \bibnamefont{Fry}},
  \bibinfo{author}{\bibfnamefont{I.~E.} \bibnamefont{Itskevich}},
  \bibinfo{author}{\bibfnamefont{D.~J.} \bibnamefont{Mowbray}},
  \bibinfo{author}{\bibfnamefont{M.~S.} \bibnamefont{Skolnick}},
  \bibinfo{author}{\bibfnamefont{J.~J.} \bibnamefont{Finley}},
  \bibinfo{author}{\bibfnamefont{J.~A.} \bibnamefont{Barker}},
  \bibinfo{author}{\bibfnamefont{E.~P.} \bibnamefont{O{'}Reilly}},
  \bibinfo{author}{\bibfnamefont{L.~R.} \bibnamefont{Wilson}},
  \bibinfo{author}{\bibfnamefont{I.~A.} \bibnamefont{Larkin}},
  \bibinfo{author}{\bibfnamefont{P.~A.} \bibnamefont{Maksym}},
  \bibnamefont{et~al.}, \bibinfo{journal}{Phys. Rev. Lett.}
  \textbf{\bibinfo{volume}{84}}, \bibinfo{pages}{733} (\bibinfo{year}{2000}).

\bibitem[{\citenamefont{Yoffe}(2001)}]{yoffe}
\bibinfo{author}{\bibfnamefont{A.~D.} \bibnamefont{Yoffe}},
  \bibinfo{journal}{Adv. Phys.} \textbf{\bibinfo{volume}{50}},
  \bibinfo{pages}{1} (\bibinfo{year}{2001}).

\bibitem[{\citenamefont{Empedocles and Bawendi}(1997)}]{science97}
\bibinfo{author}{\bibfnamefont{S.~A.} \bibnamefont{Empedocles}}
  \bibnamefont{and} \bibinfo{author}{\bibfnamefont{M.~G.}
  \bibnamefont{Bawendi}}, \bibinfo{journal}{Science}
  \textbf{\bibinfo{volume}{278}}, \bibinfo{pages}{2114} (\bibinfo{year}{1997}).

\bibitem[{\citenamefont{Blanton et~al.}(1996)\citenamefont{Blanton, Hines, and
  Guyot-Sionnest}}]{blanton}
\bibinfo{author}{\bibfnamefont{S.~A.} \bibnamefont{Blanton}},
  \bibinfo{author}{\bibfnamefont{M.~A.} \bibnamefont{Hines}}, \bibnamefont{and}
  \bibinfo{author}{\bibfnamefont{P.}~\bibnamefont{Guyot-Sionnest}},
  \bibinfo{journal}{Appl. Phys. Lett.} \textbf{\bibinfo{volume}{69}},
  \bibinfo{pages}{3905} (\bibinfo{year}{1996}).

\bibitem[{\citenamefont{Vi{\~n}a et~al.}(1987)\citenamefont{Vi{\~n}a, Collins,
  Mendez, and Wang}}]{vina87}
\bibinfo{author}{\bibfnamefont{L.}~\bibnamefont{Vi{\~n}a}},
  \bibinfo{author}{\bibfnamefont{R.~T.} \bibnamefont{Collins}},
  \bibinfo{author}{\bibfnamefont{E.~E.} \bibnamefont{Mendez}},
  \bibnamefont{and} \bibinfo{author}{\bibfnamefont{W.~I.} \bibnamefont{Wang}},
  \bibinfo{journal}{Phys. Rev. Lett.} \textbf{\bibinfo{volume}{58}},
  \bibinfo{pages}{832} (\bibinfo{year}{1987}).

\bibitem[{\citenamefont{Bauer and Ando}(1987)}]{bauer87}
\bibinfo{author}{\bibfnamefont{G.~E.~W.} \bibnamefont{Bauer}} \bibnamefont{and}
  \bibinfo{author}{\bibfnamefont{T.}~\bibnamefont{Ando}},
  \bibinfo{journal}{Phys. Rev. Lett.} \textbf{\bibinfo{volume}{59}},
  \bibinfo{pages}{601} (\bibinfo{year}{1987}).

\bibitem[{\citenamefont{Bauer and Ando}(1988)}]{bauer88b}
\bibinfo{author}{\bibfnamefont{G.~E.~W.} \bibnamefont{Bauer}} \bibnamefont{and}
  \bibinfo{author}{\bibfnamefont{T.}~\bibnamefont{Ando}},
  \bibinfo{journal}{Phys. Rev. B} \textbf{\bibinfo{volume}{38}},
  \bibinfo{pages}{6015} (\bibinfo{year}{1988}).

\bibitem[{\citenamefont{Canham}(1990)}]{canham90}
\bibinfo{author}{\bibfnamefont{L.~T.} \bibnamefont{Canham}},
  \bibinfo{journal}{Appl. Phys. Lett.} \textbf{\bibinfo{volume}{57}},
  \bibinfo{pages}{1046} (\bibinfo{year}{1990}).

\bibitem[{\citenamefont{Wolkin et~al.}(1999)\citenamefont{Wolkin, Jorne,
  Fauchet, Allan, and Delerue}}]{wolkin99}
\bibinfo{author}{\bibfnamefont{M.~V.} \bibnamefont{Wolkin}},
  \bibinfo{author}{\bibfnamefont{J.}~\bibnamefont{Jorne}},
  \bibinfo{author}{\bibfnamefont{P.~M.} \bibnamefont{Fauchet}},
  \bibinfo{author}{\bibfnamefont{G.}~\bibnamefont{Allan}}, \bibnamefont{and}
  \bibinfo{author}{\bibfnamefont{C.}~\bibnamefont{Delerue}},
  \bibinfo{journal}{Phys. Rev. Lett.} \textbf{\bibinfo{volume}{82}},
  \bibinfo{pages}{197} (\bibinfo{year}{1999}).

\bibitem[{\citenamefont{Efros et~al.}(1996)\citenamefont{Efros, Rosen, Kuno,
  Nirmal, Norris, and Bawendi}}]{rosen}
\bibinfo{author}{\bibfnamefont{A.~L.} \bibnamefont{Efros}},
  \bibinfo{author}{\bibfnamefont{M.}~\bibnamefont{Rosen}},
  \bibinfo{author}{\bibfnamefont{M.}~\bibnamefont{Kuno}},
  \bibinfo{author}{\bibfnamefont{M.}~\bibnamefont{Nirmal}},
  \bibinfo{author}{\bibfnamefont{D.~J.} \bibnamefont{Norris}},
  \bibnamefont{and} \bibinfo{author}{\bibfnamefont{M.}~\bibnamefont{Bawendi}},
  \bibinfo{journal}{Phys. Rev. B} \textbf{\bibinfo{volume}{54}},
  \bibinfo{pages}{4843} (\bibinfo{year}{1996}).

\bibitem[{\citenamefont{Casado and Trallero-Giner}(1996)}]{casadostark}
\bibinfo{author}{\bibfnamefont{E.}~\bibnamefont{Casado}} \bibnamefont{and}
  \bibinfo{author}{\bibfnamefont{C.}~\bibnamefont{Trallero-Giner}},
  \bibinfo{journal}{Phys. Status Solidi B} \textbf{\bibinfo{volume}{196}},
  \bibinfo{pages}{335} (\bibinfo{year}{1996}).

\bibitem[{\citenamefont{Wen et~al.}(1995)\citenamefont{Wen, Lin, Jiang, and
  Chen}}]{wen95}
\bibinfo{author}{\bibfnamefont{G.~W.} \bibnamefont{Wen}},
  \bibinfo{author}{\bibfnamefont{J.~Y.} \bibnamefont{Lin}},
  \bibinfo{author}{\bibfnamefont{H.~X.} \bibnamefont{Jiang}}, \bibnamefont{and}
  \bibinfo{author}{\bibfnamefont{Z.}~\bibnamefont{Chen}},
  \bibinfo{journal}{Phys. Rev. B} \textbf{\bibinfo{volume}{52}},
  \bibinfo{pages}{5913} (\bibinfo{year}{1995}).

\bibitem[{\citenamefont{Baldereschi and Lipari}(1973)}]{bald}
\bibinfo{author}{\bibfnamefont{A.}~\bibnamefont{Baldereschi}} \bibnamefont{and}
  \bibinfo{author}{\bibfnamefont{N.~O.} \bibnamefont{Lipari}},
  \bibinfo{journal}{Phys. Rev. B} \textbf{\bibinfo{volume}{8}},
  \bibinfo{pages}{2697} (\bibinfo{year}{1973}).

\bibitem[{\citenamefont{Chang and Xia}(1998)}]{baixia98}
\bibinfo{author}{\bibfnamefont{K.}~\bibnamefont{Chang}} \bibnamefont{and}
  \bibinfo{author}{\bibfnamefont{J.-B.} \bibnamefont{Xia}},
  \bibinfo{journal}{J. Appl. Phys.} \textbf{\bibinfo{volume}{84}},
  \bibinfo{pages}{1454} (\bibinfo{year}{1998}).

\bibitem[{\citenamefont{Men{\'e}ndez-Proupin}(2003)}]{comment2chang98}
\bibinfo{author}{\bibfnamefont{E.}~\bibnamefont{Men{\'e}ndez-Proupin}},
  \bibinfo{journal}{J. Appl. Phys.}  (\bibinfo{year}{2003}),
  \bibinfo{note}{submitted}.

\bibitem[{\citenamefont{Men\'{e}ndez et~al.}(1997)\citenamefont{Men\'{e}ndez,
  Trallero-Giner, and Cardona}}]{e1}
\bibinfo{author}{\bibfnamefont{E.}~\bibnamefont{Men\'{e}ndez}},
  \bibinfo{author}{\bibfnamefont{C.}~\bibnamefont{Trallero-Giner}},
  \bibnamefont{and} \bibinfo{author}{\bibfnamefont{M.}~\bibnamefont{Cardona}},
  \bibinfo{journal}{Phys. Status Solidi B} \textbf{\bibinfo{volume}{199}},
  \bibinfo{pages}{81} (\bibinfo{year}{1997}).

\bibitem[{\citenamefont{Jackson}(1962)}]{jackson}
\bibinfo{author}{\bibfnamefont{J.~D.} \bibnamefont{Jackson}},
  \emph{\bibinfo{title}{Classical Electrodynamics}}
  (\bibinfo{publisher}{Wiley}, \bibinfo{address}{New York},
  \bibinfo{year}{1962}).

\bibitem[{\citenamefont{Aggarwal}(1974)}]{aggarwal74}
\bibinfo{author}{\bibfnamefont{L.~R.} \bibnamefont{Aggarwal}}, in
  \emph{\bibinfo{booktitle}{Modulation Techniques}}, edited by
  \bibinfo{editor}{\bibfnamefont{P.~K.} \bibnamefont{Willardson}}
  \bibnamefont{and} \bibinfo{editor}{\bibfnamefont{A.~C.} \bibnamefont{Beer}}
  (\bibinfo{publisher}{Academic}, \bibinfo{address}{New York},
  \bibinfo{year}{1974}), vol.~\bibinfo{volume}{9} of
  \emph{\bibinfo{series}{Semiconductors and Semimetals}}, p.
  \bibinfo{pages}{151}.

\bibitem[{\citenamefont{Brink and Satcher}(1968)}]{brink}
\bibinfo{author}{\bibfnamefont{D.~M.} \bibnamefont{Brink}} \bibnamefont{and}
  \bibinfo{author}{\bibfnamefont{G.~R.} \bibnamefont{Satcher}},
  \emph{\bibinfo{title}{Angular Momentum}} (\bibinfo{publisher}{Clarendon
  Press}, \bibinfo{address}{Oxford}, \bibinfo{year}{1968}).

\bibitem[{\citenamefont{Laheld and Einevoll}(1997)}]{laheld}
\bibinfo{author}{\bibfnamefont{U.~E.~H.} \bibnamefont{Laheld}}
  \bibnamefont{and} \bibinfo{author}{\bibfnamefont{G.~T.}
  \bibnamefont{Einevoll}}, \bibinfo{journal}{Phys. Rev. B}
  \textbf{\bibinfo{volume}{55}}, \bibinfo{pages}{5184} (\bibinfo{year}{1997}).

\bibitem[{\citenamefont{Madelung}(1986)}]{landolt}
\bibinfo{editor}{\bibfnamefont{O.}~\bibnamefont{Madelung}}, ed.,
  \emph{\bibinfo{title}{Landolt-B\"{o}rnstein Numerical Data and Functional
  Relationships in Science and Technology}}, vol. \bibinfo{volume}{III/22}
  (\bibinfo{publisher}{Springer}, \bibinfo{address}{Berlin},
  \bibinfo{year}{1986}).

\bibitem[{\citenamefont{Hermann and Weisbuch}(1977)}]{hermann}
\bibinfo{author}{\bibfnamefont{C.}~\bibnamefont{Hermann}} \bibnamefont{and}
  \bibinfo{author}{\bibfnamefont{C.}~\bibnamefont{Weisbuch}},
  \bibinfo{journal}{Phys. Rev. B} \textbf{\bibinfo{volume}{15}},
  \bibinfo{pages}{823} (\bibinfo{year}{1977}).

\bibitem[{\citenamefont{Men\'{e}ndez-Proupin
  et~al.}(1999)\citenamefont{Men\'{e}ndez-Proupin, Trallero-Giner, and
  Garc\'{\i}a-Cristobal}}]{hraman}
\bibinfo{author}{\bibfnamefont{E.}~\bibnamefont{Men\'{e}ndez-Proupin}},
  \bibinfo{author}{\bibfnamefont{C.}~\bibnamefont{Trallero-Giner}},
  \bibnamefont{and}
  \bibinfo{author}{\bibfnamefont{A.}~\bibnamefont{Garc\'{\i}a-Cristobal}},
  \bibinfo{journal}{Phys. Rev. B} \textbf{\bibinfo{volume}{60}},
  \bibinfo{pages}{5513} (\bibinfo{year}{1999}).

\bibitem[{\citenamefont{Xia}(1989)}]{baixia}
\bibinfo{author}{\bibfnamefont{J.-B.} \bibnamefont{Xia}},
  \bibinfo{journal}{Phys. Rev. B} \textbf{\bibinfo{volume}{40}},
  \bibinfo{pages}{8500} (\bibinfo{year}{1989}).

\bibitem[{\citenamefont{Efros}(1992)}]{efros92}
\bibinfo{author}{\bibfnamefont{A.~L.} \bibnamefont{Efros}},
  \bibinfo{journal}{Phys. Rev. B} \textbf{\bibinfo{volume}{46}},
  \bibinfo{pages}{7448} (\bibinfo{year}{1992}).

\bibitem[{\citenamefont{Baldereschi and Lipari}(1974)}]{bald74}
\bibinfo{author}{\bibfnamefont{A.}~\bibnamefont{Baldereschi}} \bibnamefont{and}
  \bibinfo{author}{\bibfnamefont{N.~O.} \bibnamefont{Lipari}},
  \bibinfo{journal}{Phys. Rev. B} \textbf{\bibinfo{volume}{9}},
  \bibinfo{pages}{1525} (\bibinfo{year}{1974}).

\bibitem[{\citenamefont{Koster et~al.}(1963)\citenamefont{Koster, Dimmock,
  Wheeler, and Statz}}]{koster63}
\bibinfo{author}{\bibfnamefont{G.~F.} \bibnamefont{Koster}},
  \bibinfo{author}{\bibfnamefont{J.~O.} \bibnamefont{Dimmock}},
  \bibinfo{author}{\bibfnamefont{R.~G.} \bibnamefont{Wheeler}},
  \bibnamefont{and} \bibinfo{author}{\bibfnamefont{H.}~\bibnamefont{Statz}},
  \emph{\bibinfo{title}{Properties of the thirty-two point groups}}
  (\bibinfo{publisher}{M.I.T. Press}, \bibinfo{address}{Cambridge,
  Massachusetts}, \bibinfo{year}{1963}).

\bibitem[{\citenamefont{Men\'{e}ndez-Proupin and
  Cabo-Bisset}(2002)}]{articlehr}
\bibinfo{author}{\bibfnamefont{E.}~\bibnamefont{Men\'{e}ndez-Proupin}}
  \bibnamefont{and}
  \bibinfo{author}{\bibfnamefont{N.}~\bibnamefont{Cabo-Bisset}},
  \bibinfo{journal}{Phys. Rev. B} \textbf{\bibinfo{volume}{66}},
  \bibinfo{pages}{085317} (\bibinfo{year}{2002}).

\bibitem[{\citenamefont{Kash et~al.}(1985)\citenamefont{Kash, Mendez, and
  Morko{\c c}}}]{Kash}
\bibinfo{author}{\bibfnamefont{J.~A.} \bibnamefont{Kash}},
  \bibinfo{author}{\bibfnamefont{E.~E.} \bibnamefont{Mendez}},
  \bibnamefont{and} \bibinfo{author}{\bibfnamefont{H.}~\bibnamefont{Morko{\c
  c}}}, \bibinfo{journal}{Appl. Phys. Lett.} \textbf{\bibinfo{volume}{46}},
  \bibinfo{pages}{173} (\bibinfo{year}{1985}).

\bibitem[{\citenamefont{Miller and Gossard}(1983)}]{Miller83}
\bibinfo{author}{\bibfnamefont{R.~C.} \bibnamefont{Miller}} \bibnamefont{and}
  \bibinfo{author}{\bibfnamefont{A.~C.} \bibnamefont{Gossard}},
  \bibinfo{journal}{Appl. Phys. Lett.} \textbf{\bibinfo{volume}{43}},
  \bibinfo{pages}{954} (\bibinfo{year}{1983}).

\bibitem[{\citenamefont{Wang}(2001)}]{wang01}
\bibinfo{author}{\bibfnamefont{L.-W.} \bibnamefont{Wang}}, \bibinfo{journal}{J.
  Phys. Chem. B} \textbf{\bibinfo{volume}{105}}, \bibinfo{pages}{2360}
  (\bibinfo{year}{2001}).

\bibitem[{\citenamefont{Eychm{\"u}ller}(2000)}]{eychm00}
\bibinfo{author}{\bibfnamefont{A.}~\bibnamefont{Eychm{\"u}ller}},
  \bibinfo{journal}{J. Phys. Chem. B} \textbf{\bibinfo{volume}{104}},
  \bibinfo{pages}{6514} (\bibinfo{year}{2000}).

\bibitem[{\citenamefont{Tittel et~al.}(1997)\citenamefont{Tittel, G{\"o}hde,
  Koberling, Basch\'e, Kornowski, Weller, and Eychm{\"u}ller}}]{title97}
\bibinfo{author}{\bibfnamefont{J.}~\bibnamefont{Tittel}},
  \bibinfo{author}{\bibfnamefont{W.}~\bibnamefont{G{\"o}hde}},
  \bibinfo{author}{\bibfnamefont{F.}~\bibnamefont{Koberling}},
  \bibinfo{author}{\bibfnamefont{T.}~\bibnamefont{Basch\'e}},
  \bibinfo{author}{\bibfnamefont{A.}~\bibnamefont{Kornowski}},
  \bibinfo{author}{\bibfnamefont{H.}~\bibnamefont{Weller}}, \bibnamefont{and}
  \bibinfo{author}{\bibfnamefont{A.}~\bibnamefont{Eychm{\"u}ller}},
  \bibinfo{journal}{J. Phys. Chem. B} \textbf{\bibinfo{volume}{101}},
  \bibinfo{pages}{3013} (\bibinfo{year}{1997}).

\bibitem[{\citenamefont{Shim et~al.}(2001)\citenamefont{Shim, Wang, and
  Guyot-Sionnest}}]{shim01}
\bibinfo{author}{\bibfnamefont{M.}~\bibnamefont{Shim}},
  \bibinfo{author}{\bibfnamefont{C.}~\bibnamefont{Wang}}, \bibnamefont{and}
  \bibinfo{author}{\bibfnamefont{P.}~\bibnamefont{Guyot-Sionnest}},
  \bibinfo{journal}{J. Phys. Chem. B} \textbf{\bibinfo{volume}{105}},
  \bibinfo{pages}{2369} (\bibinfo{year}{2001}).

\bibitem[{\citenamefont{Rabani et~al.}(1999)\citenamefont{Rabani, Het\'{e}nyi,
  Berne, and Brus}}]{rabani99}
\bibinfo{author}{\bibfnamefont{E.}~\bibnamefont{Rabani}},
  \bibinfo{author}{\bibfnamefont{B.}~\bibnamefont{Het\'{e}nyi}},
  \bibinfo{author}{\bibfnamefont{B.~J.} \bibnamefont{Berne}}, \bibnamefont{and}
  \bibinfo{author}{\bibfnamefont{L.~E.} \bibnamefont{Brus}},
  \bibinfo{journal}{J. Chem. Phys.} \textbf{\bibinfo{volume}{110}},
  \bibinfo{pages}{5355} (\bibinfo{year}{1999}).

\bibitem[{\citenamefont{Fu}(2002)}]{fu02}
\bibinfo{author}{\bibfnamefont{H.}~\bibnamefont{Fu}}, \bibinfo{journal}{Phys.
  Rev. B} \textbf{\bibinfo{volume}{65}}, \bibinfo{pages}{045320}
  (\bibinfo{year}{2002}).

\bibitem[{\citenamefont{Fomin et~al.}(1998)\citenamefont{Fomin, Gladilin,
  Devreese, Pokatilov, Balaban, and Klimin}}]{fomin}
\bibinfo{author}{\bibfnamefont{V.~M.} \bibnamefont{Fomin}},
  \bibinfo{author}{\bibfnamefont{V.~N.} \bibnamefont{Gladilin}},
  \bibinfo{author}{\bibfnamefont{J.~T.} \bibnamefont{Devreese}},
  \bibinfo{author}{\bibfnamefont{E.~P.} \bibnamefont{Pokatilov}},
  \bibinfo{author}{\bibfnamefont{S.~N.} \bibnamefont{Balaban}},
  \bibnamefont{and} \bibinfo{author}{\bibfnamefont{S.~N.}
  \bibnamefont{Klimin}}, \bibinfo{journal}{Phys. Rev. B}
  \textbf{\bibinfo{volume}{57}}, \bibinfo{pages}{2415} (\bibinfo{year}{1998}).

\bibitem[{\citenamefont{Banyai et~al.}(1992)\citenamefont{Banyai, Gilliot, Hu,
  and Koch}}]{banyai92}
\bibinfo{author}{\bibfnamefont{L.}~\bibnamefont{Banyai}},
  \bibinfo{author}{\bibfnamefont{P.}~\bibnamefont{Gilliot}},
  \bibinfo{author}{\bibfnamefont{Y.~Z.} \bibnamefont{Hu}}, \bibnamefont{and}
  \bibinfo{author}{\bibfnamefont{S.~W.} \bibnamefont{Koch}},
  \bibinfo{journal}{Phys. Rev. B} \textbf{\bibinfo{volume}{45}},
  \bibinfo{pages}{14136} (\bibinfo{year}{1992}).

\end{thebibliography}

\end{document}